\documentclass[letterpaper,twocolumn,10pt]{article}
\usepackage{format}

\usepackage{booktabs}
\usepackage{graphicx}
\usepackage[numbers]{natbib}
\usepackage[absolute]{textpos}
\usepackage{hyperref}

\begin{document}

\date{}

\title{\Large \bf Trustworthy Transparency by Design}

\def\plainauthor{Valentin Zieglmeier and Alexander Pretschner}

\newcommand{\vzinst}{Technical University of Munich, Germany}

\author{
{\rm Valentin Zieglmeier}\\
\vzinst{}\\
valentin.zieglmeier@tum.de
\and
{\rm Alexander Pretschner}\\
\vzinst{}\\
alexander.pretschner@tum.de
} 

\maketitle
\thecopyright

\begin{textblock*}{.75\paperwidth}(.125\paperwidth,.07\paperwidth)
	\noindent\framebox[\textwidth]{ 
		\parbox{\dimexpr\linewidth-2\fboxsep-2\fboxrule}{%
			\centering \textbf{Note:} This paper has been significantly revised, with new studies and a more robust theoretical background. The new version is at \href{https://arxiv.org/abs/2305.09813}{arXiv:2305.09813}. It differs substantially, though, so we keep this version online for reference.
		}
	}
\end{textblock*}

\newcommand{\citeauthorcite}[1]{\citeauthor{#1}~\cite{#1}}

\hyphenation{tool-chain}

\begin{abstract}
Individuals lack oversight over systems that process their data. This can lead to discrimination and hidden biases that are hard to uncover.
Recent data protection legislation tries to tackle these issues, but it is inadequate.
It does not prevent data misusage while stifling sensible use cases for data.

We think the conflict between data protection and increasingly data-based systems should be solved differently. When access to data is given, all usages should be made transparent to the data subjects.
This enables their \emph{data sovereignty}, allowing individuals to benefit from sensible data usage while addressing potentially harmful consequences of data misusage.

We contribute to this with a technical concept and an empirical evaluation.
First, we conceptualize a transparency framework for software design, incorporating research on user trust and experience.
Second, we instantiate and empirically evaluate the framework in a focus group study over three months, centering on the user perspective.

Our transparency framework enables developing software that incorporates transparency in its design.
The evaluation shows that it satisfies usability and trustworthiness requirements.
The provided transparency is experienced as beneficial and participants feel empowered by it.
This shows that our framework enables \emph{Trustworthy Transparency by Design}.
\end{abstract}

\section{Introduction}

In our increasingly digitalized world, we are all dependent on technical systems that process our data.
Everything from human resources to voter registrations is now handled by computer systems, which means that data are utilized.
More often than not, we as individual users do not have control over these systems or oversight how our data are used.
In the private sector, the traditional approach of many businesses is to force users to either accept their terms of service or forfeit access to their services.
For governmental institutions on the other hand, completely opting out can often be impossible.

This black box nature of data processing raises concerns.
The lack of control and oversight means there is little recourse in case of data misusage.
With increasingly automated and automatic decision making, the risks of data misusage rise further~\cite{zhang2018mitigating, bird2019fairness, lustig2016algorithmic}.
Data-based insights may play a role in deciding if a person should be invited for a job interview, if they should get a driving license, or qualify for a loan.
It therefore becomes vital that any discrimination or misusage of data can be uncovered and challenged~\cite{rieke2018public}.

To protect individuals' privacy and ensure accountability, data protection legislation is employed~\cite{rubinstein2010privacy, flannery2017gdpr}.
Depending on the cultural context and underlying trust model, it takes different shape~\cite{bowie2006privacy, pretschner2014achieving}.
A traditional approach is to depend on self-regulation and voluntary codes, as is common in many parts of the United States of America~\cite{rubinstein2010privacy}.
This optimistic solution can be classified as \emph{detective enforcement}.
It assumes many data usages to be benign and enables them by default.
Terms of use or non-disclosure agreements enable \emph{ex post} reaction to misusage~\cite{povey1999optimistic, pretschner2014achieving}.
Recent privacy legislation such as the 2016 General Data Protection Regulation (GDPR)~\cite{eu2016gdpr} of the European Union and the 2018 California Consumer Privacy Act (CCPA)~\cite{cali2018privacy} go beyond voluntary codes to implement formal privacy regulation and provide individuals more control over their data~\cite{bowie2006privacy, rubinstein2010privacy}.
This can be seen as a move towards \emph{preventive enforcement}.
It only permits data usage for purposes specified in advance and therefore strives to prevent misusage \emph{ex ante}~\cite{vanbeek2007comparison, pretschner2008usage}.

We think this added protection is important, but it is in many cases insufficient to prevent data misusage while at the same time stifling sensible use cases for data.
Individuals are required to agree in advance to potential data usages, meaning they have to understand the impacts of giving access to their data.
This can already be a challenge when it comes to the intended usage of data, as the lengths of typical privacy policies show\footnote{Google's for example, when viewed as a PDF, is 30 pages long: \url{https://www.gstatic.com/policies/privacy/pdf/20200331/acec359e/google_privacy_policy_en_eu.pdf}} (see also~\cite{mcdonald2008cost}).
More problematically though, data that are given away may be used in \emph{unexpected} and \emph{unintended} ways~\cite{rao2016expecting, hummel2018sovereignty}, and hence be misused from the perspective of the data subject. This can happen intentionally, by trickery or hiding of essential information, or unintentionally, by misreading or misrepresenting those data.
Finally, there is little recourse in cases where data are already public or can be indirectly inferred~\cite{weitzner2008information, mundie2014privacy}.
Overwhelmed by choice and a lack of oversight, and backed by laws such as the GDPR and CCPA, individuals might therefore aim for absolute data minimization.
This ideal certainly reduces the potential for misusage, but opting out could lead to other disadvantages or might not be possible at all.
Furthermore, the \emph{preventive} approach of these laws makes it difficult to utilize data beyond cases in which they are absolutely necessary, restraining legitimately helpful data usages and stifling research and innovation in the big data space~\cite{mundie2014privacy, jia2018effects, zarsky2017incompatible, gal2020competitive}.

We think that this issue can be solved differently.
Inspired by \citeauthorcite{brin1998transparent}, we envision making all usages of individuals' data visible (transparent) to them.
Having direct oversight over how their data are used empowers individuals to gain true \emph{data sovereignty}~\cite{hummel2018sovereignty}, meaning ``self-de\-ter\-mi\-na\-tion [...] with regard to the use of their data''~\cite[p.~550]{jarke2019data}.
This has the potential to improve their trust to allow data usages that might be beneficial to them.
In addition to helping to uncover misusage retroactively which enables \emph{accountability}~\cite{weitzner2008information}, this transparency could also increase the \emph{felt accountability} of data consumers~\cite{hall2017accountability}, thereby deterring data misusage even before it occurs.

Some data, such as health or genetic data, will always warrant a \emph{preventive} approach due to their high sensitivity.
Furthermore, in scenarios where a power asymmetry is likely, such as between employer and employee~\cite[p.~166]{cas2011ubiquitous}, the added transparency alone is not sufficient to protect individuals.
Additional safeguards, such as strong workers' councils or appropriate recourse in case of data misusage (see, e.g., \cite[p.~36]{mundie2014privacy}), may therefore be a prerequisite for this idea.
Given those, we think it could be a promising solution to the conflict between data protection and data-based use cases.

To illustrate our vision, consider the example of software developers in an IT company.
These employees are in high demand in the labor market and can work remotely, enabling labor mobility.
Therefore, the inherent power asymmetry between them and their employer is balanced out.
Data about these employees are stored in various systems and accessed through a multitude of tools.
In this scenario, employees track their work in issue tracking software, which means that data about the specific technologies and problems they work on, as well as whom they collaborate with, exist.
The traditional \emph{detective} approach allows utilizing these data for, e.g., company-level decision making or collaboration between colleagues.
However, it makes room for profiling and patronization of employees based on data that might not represent the full picture or be inadequate for these uses.
Employees might be fired or discriminated against due to misinterpreted or misused data, and have no recourse against it.
With a \emph{preventive} approach on the other hand, any data usage beyond those required by basic work processes is forbidden.
This makes it difficult to implement systems enabling advanced data-based use cases.
Yet, as we have deliberated above, misusage of data is not sufficiently prevented.
If we now imagine the same example with the envisioned transparency over data usages, those issues are addressed.
Employees are free to collaborate without any overhead, and data can be utilized for company-level decision making.
Should data be misused and harmful consequences for an employee arise, they have access to a tamper-proof audit trail.
To fight back, they can make it available to their workers' council or a lawyer to support their case.


In this paper, we explore the relationship of trust and transparency to understand to which extent the \emph{data sovereignty} of individuals can be enabled with trustworthy transparency into data usage processes.
Our goal is to facilitate data-based use cases that can be beneficial for individuals while protecting them from misusage of their data.
To that end, we describe a technical concept for a transparency framework that enables \emph{Transparency by Design}.
It is based on the idea that if data are used, those usages are tracked and made transparent to data subjects.
In order for the framework to be useful, it depends on the acceptance and trust of individuals.
Therefore, we instantiate and evaluate the framework to understand the user perspective on our concept.
Our evaluation is designed as a focus group study and mirrors a real-world use case provided by our industry partner.
We assess and discuss the user experience and personal perspectives of participants, as well as the perceived trustworthiness of the transparency framework.


The idea of providing individuals with transparency about how their data are used has existed for a long time (see Section~\ref{sec:related-work}).
Many previous works are conceptual (e.g., \cite{brin1998transparent, agrawal2002hippocratic, weitzner2008information, dabrock2019data}), serving as a foundation to our work.
We implement their idea in practice and show its effect on individuals.
Some go in different directions, which does not sufficiently address the problems we have outlined or introduces new issues that we want to avoid (e.g., \cite{gates2003owner, demontjoye2014openpds, papadopoulou2015enabling, birrell2018sgx}; see related work).
For example, some authors propose having individuals store all of their data themselves, introducing problems with data security and validation~\cite{gates2003owner}.
Finally, there are proposals that go in the same direction:
The \emph{CIA framework} tries to enable usage logging for data based on Java JAR files~\cite{sundareswaran2011promoting}.
The privacy platform \emph{Ancile} automatically audits Python scripts sent to its server~\cite{bagdasaryan2019ancile}.
These existing implementations do not cover two important aspects:
First, they are use case specific and therefore cannot be generalized to enable transparency into any data usage process.
Second, they do not consider the user perspective, which is an important factor to empower individuals and enable their acceptance and trust.
For a full discussion of related work, see Section~\ref{sec:related-work}.


We contribute to closing these gaps with a technical concept and an empirical evaluation.
Our concept for a transparency framework allows integrating \emph{Transparency by Design} into software systems.
Based on research on user trust and experience, we describe how we systematically foster trust in the framework and improve its user experience.
In our evaluation, we find that our framework satisfies usability and trustworthiness requirements.
Finally, we present evidence that participants of our focus group find the transparency they experienced beneficial and feel empowered by it.

\section{Related Work}
\label{sec:related-work}

The concept of \emph{data sovereignty} and its application to strengthen individuals' data privacy has been developed by \citeauthorcite{hummel2018sovereignty} and \citeauthorcite{dabrock2019data}. The latter in particular discusses a technical concept that shares many similarities with our approach, mainly in that a so-called \emph{data agent} tracks transmission and processing of data~\cite{dabrock2019data}. Contrary to their idea of automatically deciding which data usage is acceptable and which is not, we focus on making transparent the logged usages to data owners.
\Citeauthor{dabrock2019data} does not consider the possibility of \emph{unexpected} and \emph{unintended} usages of data that only become apparent through transparency into data usage processes.
There is a gray area of usages that might be technically allowed, but questionable.
With our approach of full transparency into data usage processes, we think we can enable the \emph{(felt) accountability} of data consumers~\cite{weitzner2008information, hall2017accountability}.
In addition, this direct oversight is in our view a prerequisite for true \emph{data sovereignty} of individuals.

There have been many great thinkers who have contributed to shaping this thought.
We are inspired by \citeauthorcite{brin1998transparent}, who first described the idea of allowing full transparency over any data usage.
Another important predecessor to our work is the paper on \emph{hippocratic databases} by \citeauthorcite{agrawal2002hippocratic}. They discuss the usefulness of giving individuals access to audit trails of databases holding their information, allowing them to detect misusage~\cite{agrawal2002hippocratic}.
Finally, \citeauthor{weitzner2008information} also describe the potential of making data usages transparent to individuals, achieving what they refer to as \emph{information accountability}~\cite{weitzner2008information}.
They see two main advantages over the status quo:
First, reducing individuals' mental load as they do not have to judge \emph{ex ante} all potential usages of their data.
Second, enabling redress in case of harmful misusage of data~\cite{weitzner2008information}.
These works serve as a theoretical foundation to our concept.
We implement their idea in practice and show its effect on individuals.

\citeauthorcite{gates2003owner} take our concept a step further to maximize the \emph{data sovereignty} of individuals. They envision having individuals store all of their data themselves, retaining full control. There exist multiple other concepts that aim in a similar direction (e.g., \cite{hong2004architecture, mun2010personal, demontjoye2014openpds, papadopoulou2015enabling}). We refer to this idea as \emph{personal data stores}.\footnote{Term coined by \citeauthorcite{demontjoye2014openpds} and \citeauthorcite{papadopoulou2015enabling}.}
The overarching goal is sensible, but we consider the approach of creating \emph{personal data stores} non-ideal. It introduces multiple challenges that do not seem to outweigh the benefits, primarily the increased responsibility laid on individuals to keep their data secure~\cite{gates2003owner}.
Another important problem is data authenticity or provenance, especially critical for data that the owner might be interested to manipulate~\cite{gates2003owner, papadopoulou2015enabling}.
Our approach gives individuals similar power.
At the same time, they are not burdened with problems related to having to safekeep their data themselves.
Furthermore, we avoid the problem of data validation for data consumers.

Finally, \citeauthorcite{sundareswaran2011promoting} and \citeauthorcite{bagdasaryan2019ancile} describe technical concepts that share many ideas with ours.
The \emph{CIA framework}~\cite{sundareswaran2011promoting} tries to solve a similar problem to ours on a very technical level.
The idea of incorporating logging based on the behavior of the app is implemented with Java JAR files~\cite{sundareswaran2011promoting}.
\emph{Ancile}~\cite{bagdasaryan2019ancile} on the other hand is an online privacy platform. It also identifies many of the same challenges that we see.
This technical implementation considers Python-based analysis scripts that are sent to the \emph{Ancile} server for execution.
Both concepts are very promising and relevant, but they both do not address two important aspects:
First, their implementations are highly use case specific, which makes them unfit for generalization. The JAR-based logging of the \emph{CIA framework} for example can only be implemented in programs that can handle data distributed as binaries.
We go beyond these technical proposals to describe a methodology of incorporating transparency into any software design.
Second, both do not consider the user perspective, a vital aspect of privacy and empowerment.
Based on the state of the art in user trust research, we conceptualize factors that can foster usability and trust.
We apply those in an instantiation of our framework and evaluate it with a focus on user acceptance and trust.
In our focus group study, we assess if participants consider the experienced transparency to be beneficial and feel empowered by it.
Finally, we deliberate the factors that contribute to this effect.

\section{Background: Trust and Transparency}
\label{sec:transparency-trust}

Before describing the envisioned transparency framework, we first give an overview over research on \emph{trust} and \emph{transparency}, their relationship to each other, as well as their relevance for our context.

User trust in technology is important, considering that it influences users' intention to use, adoption, and continued use~\cite{riegelsberger2005mechanics, komiak2006effects, sollner2012understanding}. Furthermore, trust antecedents such as credibility can increase intention to use as well~\cite{ong2004factors}.

Research on trust can be differentiated between \emph{interpersonal trust} and \emph{trust in automation}.
For interpersonal trust, the formative model conceived by \citeauthor{mayer1995integrative} defines three central antecedents for trust: \emph{ability}, \emph{integrity}, and \emph{benevolence}~\cite{mayer1995integrative}.
Corresponding to these, for trust in automation the antecedents of \emph{purpose}, \emph{process}, and \emph{performance} were defined by \citeauthorcite{lee2004trust} based on earlier work by \citeauthorcite{lee1992trust}.
Their model has been shown to better reflect users' beliefs in the context of IT artifacts~\cite{sollner2012understanding}.
It is commonly applied in research on user trust in the context of various technological applications (see, e.g., \cite{hoffmann2014incorporating, hoff2015trust, chien2018effect}).
User trust and trust antecedents have been researched for different application areas of computer science, such as software tools~\cite{vila2011consumer}, automation~\cite{lee2004trust}, algorithms~\cite{kizilcec2016much}, or context-aware systems~\cite{antifakos2005towards}, among various others.

Transparency itself can be a trust antecedent~\cite{hoff2015trust, hancock2011can, schnorf2014trust}. In the context of technology, providing transparency means informing the user about the system's function (purpose) and functionality (process)~\cite{hancock2011can, hoff2015trust}.
The influence of transparency on trust has been shown relating to how software tools are designed on a higher level~\cite{schnorf2014trust, constantine2006trusted, cramer2007user} but also for the underlying algorithms~\cite{kizilcec2016much, ananny2018seeing, lepri2018fair}.

There are limits as to how far transparency is effective in fostering trust, depending on how it is provided and to which degree~\cite{kizilcec2016much, kroll2016accountable, rader2018explanations}.
Too much transparency~\cite{kizilcec2016much} or ``wrong'' ways to frame this transparency~\cite{rader2018explanations} may even reduce trust.
In addition, if explanations or insights are too technical, this too can render them ineffective or counterproductive towards trust~\cite{rawlins1994measuring, cramer2007user, kroll2016accountable}.

Subsequently, the transparency \emph{provided} by software tools needs to also be trustworthy to be effective.
Of course, this includes ensuring technically that protection systems cannot be circumvented and, if possible, giving guarantees of that~\cite{chase2016transparency}.
That is not sufficient by itself, though. The way information is presented, framed, and the human factors that surround a system's implementation or rollout are of equal importance~\cite{hoffmann2013fostering, faisal2016web, fisher2020does}.
To illustrate this: If a system is built to be cryptographically secure, but users are not informed of this fact or do not trust the messenger, this technical fact alone does little to improve their trust in the system.

\section{Enabling Trustworthy Transparency Into Data Usage Processes}

Currently, data usage processes happen without oversight of the individuals concerned by them.
For example, people analytics software is used to understand and steer the workforce in many companies.
These tools access a variety of data, such as information about employees' conduct or performance.
If data are analyzed or interpreted through these tools, this is nontransparent to the data subjects~\cite{arellano2017using, gal2020breaking}.
The insights gained from these tools may disadvantage or discriminate, but this can be almost impossible for individuals to uncover~\cite{zarsky2016trouble}.
There might be documents describing in theory how data may be used, such as company agreements or terms of service. Yet, it is all but impossible for an individual to oversee that these agreements are followed in practice.
In addition, both intentionally and unintentionally, data that have been given away can be used in ways that are \emph{unexpected} and \emph{unintended} to individuals, something that we consider misusage of data.

To tackle this, various authors have proposed making data usage processes transparent to the data subjects (e.g.,~\cite{agrawal2002hippocratic, weitzner2008information}).
Based on their conceptual work, we derive a technical concept for a transparency framework.
Our overarching vision is to devise an entirely new software design. Following \emph{Privacy by Design}~\cite{cavoukian2009privacy}, we envision \emph{Transparency by Design}. All software should be built in such a way that data usages can be limited, prevented, and---most importantly---traced back.
Our transparency framework concept represents a software design pattern that enables this \emph{Transparency by Design}.

Previous works towards this goal do not cover two important aspects:
First, they omit the global perspective.
The solutions are often highly use case specific (see, e.g., \cite{sundareswaran2011promoting}) and therefore cannot be generalized to enable transparency into any data usage process.
Our framework describes a high level pattern. Depending on the use case, previous works can be integrated into our framework to enable monitoring of specific applications (see Section~\ref{sec:concept-related-work}).
Second, they omit the user perspective.
To foster user trust and acceptance, it is important to understand the user perspective on the provided transparency.
This is a prerequisite to empower individuals and enable their \emph{data sovereignty}.
Therefore, we describe how we systematically foster trust in the framework and improve the user experience of its instantiation.
We derive these steps from the state of the art in user trust and experience research (see Section~\ref{sec:concept-trustworthiness}).
Finally, we evaluate if we manage to fulfill usability and trustworthiness requirements with our framework.
To that end, we conduct a qualitative focus group study.
Our evaluation spans a three month period and closely mirrors a real-world use case provided by our industry partner.

From here on, instead of talking specifically about individuals, businesses, or governmental institutions, we will refer to the more generic concepts of \emph{data owners} and \emph{data consumers}. As defined by \citeauthor{pretschner2006distributed}, for each datum there is a \emph{data owner}~\cite{pretschner2006distributed}. They ``[possess] the rights to the data''~\cite[p.~40]{pretschner2006distributed}. Often, this corresponds to the person who created the datum.\footnotemark{} They also define the role of the \emph{data consumer}~\cite{pretschner2006distributed}.
In our concept, we personify the data consumer. Our framework can be applied to algorithmic data usages as well. In those cases, the system is considered the \emph{data consumer}.
\footnotetext{This simplification is meant to be illustrative, not conclusive. We do not consider the legal, nor the psychological or sociological dimensions of \emph{data ownership} here.}

\subsection{Concept for a Transparency Framework}
\label{sec:concept}

The idea of the transparency framework is to give data owners direct oversight over data usages. That requires (1) monitoring every usage of data, (2) verifying the authenticity of these events and storing them, and (3) making this information transparent to data owners. In its basic form, the components of the framework can be imagined as individual tools or services, but they can also be integrated into the operating system or directly into the software tools that provide data access.

To enable the three steps of our vision, we conceive three (conceptual) components:

\begin{enumerate}
	\item \emph{Monitor}: Track data usages
	\item \emph{Safekeeper}: Verify and store monitored usages
	\item \emph{Display}: Make stored usages transparent
\end{enumerate}

In the following, we describe how these components work together by discussing two aspects: Their functionality---how they actually make data usages transparent---as well as how to systematically foster their trustworthiness.

\subsubsection{Functionality}

In order to make all data usages transparent, we need to not only track all occurring usages, but also prevent circumvention of the framework logging. Therefore, it is important to ensure that data never leave controlled environments. Accessing them may only be allowed through tools that integrate the transparency framework, thereby ensuring that usages are logged~\cite{weitzner2008information}.
This is why we envision \emph{Transparency by Design}: No software tool should be distributed without enabling this transparency.
This requires software developers to integrate transparency into their architectures (see Sections~\ref{sec:instantiation-jira} and \ref{sec:instantiation-standalone}), and administrators to embed these tools into their infrastructures (see Section~\ref{sec:instantiation-technical-overview}).
Undoubtedly, there will always remain a risk of intentionally malicious applications. In the most sensitive environments, application auditing or signing can therefore be employed to ensure compliance.

Of course, as soon as data are provided to data consumers, even within a monitored environment, our control over it ends. It is nigh impossible to prevent them from simply taking a picture of the screen~\cite{pretschner2008usage} or even just memorizing its contents~\cite{pretschner2013distributed}.
Still, we argue that this does not significantly reduce the value of the provided transparency.
Large-scale copying or exporting of data is not feasible this way, making it only a theoretical issue.
Therefore, in reality, we expect all data usages to go through the framework, making them traceable.

\begin{figure}[htbp]
	\centering
	\includegraphics[width=0.9\linewidth]{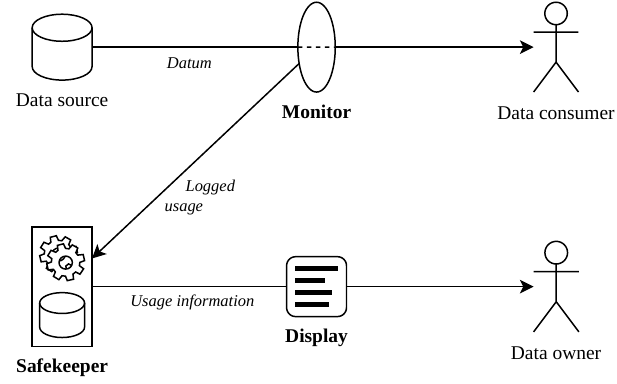}
	\caption{Making data usages transparent through the transparency framework. The data consumer accesses a datum from the data source. Their usage is logged and stored. The data owner can now retrieve the logged usage information.}
	\label{fig:framework-concept}
\end{figure}

Figure~\ref{fig:framework-concept} shows the basic conceptual steps to make usages transparent through the framework.
Access to a data source is watched by the \emph{Monitor}, which logs usages to the \emph{Safekeeper} component. The data owner accesses stored usage information through the \emph{Display}.

\subsubsection{Trustworthiness}
\label{sec:concept-trustworthiness}

From a technocratic worldview, it might seem as though the only relevant factor to concern us to enable user trust would be the functionality discussed above. After all, as long as everything is logged in a tamper-proof way, our work should be complete.
This is not at all the case, though. As we have discussed in Section~\ref{sec:transparency-trust}, user trust in technology depends on a variety of factors, categorized in the dimensions of \emph{purpose}, \emph{process}, and \emph{performance}. Accordingly, we need to consider those factors to systematically achieve trustworthiness.

Aspects relating to the \emph{purpose} dimension depend on users' personal perception of the humans involved~\cite{lee2004trust, hoffmann2014incorporating}, such as the system developers or its operators.
More concretely, this can mean how clearly the designers' motives are communicated, whether users perceive the artifact to be created with benevolence towards them, and their belief that they can rely upon the artifact in the future~\cite[p.~3]{sollner2011towards}.
To improve trust in this dimension, a reputable or well-known (third) party can be made responsible for the development and operation of the transparency-enabling systems or certify their correctness, thereby targeting the trust antecedents of reputation and familiarity~\cite{hoff2015trust, chen2007initial, du2009empirical}.
Alternatively, the use of open source software or code audits can reduce dependence on interpersonal trust~\cite{rana2004trust, garcia2005trust, alarcon2020trust}.

Secondly, trust antecedents relating to \emph{performance}, such as design and aesthetics~\cite{hassenzahl2008aesthetics, yuksel2017brains, roy2001impact} as well as content display~\cite{hartmann2008framing, faisal2016web, everard2005presentation}, depend on the specific instantiation.
We will detail how we systematically improve them in Section~\ref{sec:instantiation-trustworthiness}.

Finally, the \emph{process} factors of perceived security and authenticity of displayed information~\cite{yenisey2005perceived, flavian2006consumer, chellappa2008consumers, mirnig2014trust} can be addressed on a conceptual level.
In the following, we therefore integrate them into our concept to inform the design of instantiations based on it.

These trust antecedents are related to how securely designed users perceive the system to be, necessitating reliable and trustworthy authentication and security to be built into instantiations of the framework. We argue that this can best be enabled by incorporating existing authentication services.
Utilizing established authentication infrastructure reduces friction when using the transparency framework.
More importantly, it can improve user trust: As users know the authentication system already, the trust antecedents reputation and familiarity~\cite{hoff2015trust, chen2007initial, du2009empirical} are already established. That means they might be more inclined to trust the framework as well~\cite{komiak2006effects, holzinger2011effect}. Of course, this requires the utilized authentication infrastructure itself to be trusted by users.

\begin{figure}[htbp]
	\centering
	\includegraphics[width=0.9\linewidth]{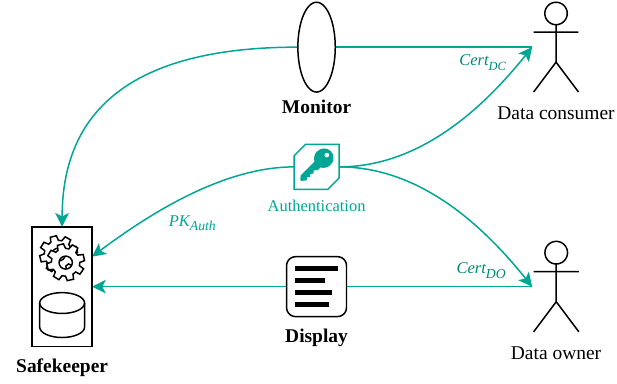}
	\caption{Integrating an existing authentication server into the framework to enable authenticity and security. The user certificates $Cert_{DC}$ and $Cert_{DO}$ are used for logging in. Now, they can be passed on by the \emph{Monitor} and \emph{Display} components. The \emph{Safekeeper} component uses the public key $PK_{Auth}$ of the \emph{Authentication} server together with the user certificates to verify the authenticity of the data it receives.}
	\label{fig:framework-architecture-concept-auth}
\end{figure}

In Figure~\ref{fig:framework-architecture-concept-auth}, the authentication flow based on an existing authentication server is shown.
This could be, for example, an access delegation service~\cite{rfc6749oauth} such as a social log-in for online services. In the context of internal company infrastructures, their single sign-on server~\cite{satoh2004single} can be utilized.
As ordinary user certificates are repurposed, no new trusted infrastructure has to be introduced. Both data owners and data consumers log in with their existing credentials. In the background, both the \emph{Monitor} and the \emph{Display} components forward the user certificates to the \emph{Safekeeper} component. Together with the public key of the authentication server, the \emph{Safekeeper} can now verify the authenticity of the data it receives~\cite{rfc7519jwt, josephson2004peer, satoh2004single}.

\subsubsection{Incorporating Related Work}
\label{sec:concept-related-work}

Our transparency framework is a high-level pattern for designing software systems. To embed it with previous work on enabling transparency into data usages, we detail how those works relate to our concept.

Various authors have proposed use case specific ways to track or limit data usage in different technical systems (e.g., \cite{sundareswaran2011promoting, bagdasaryan2019ancile, kelbert2018data, birrell2018sgx, wagner2018distributed}). For example, \citeauthor{sundareswaran2011promoting} describe how to incorporate logging into data packages distributed as Java JAR binaries~\cite{sundareswaran2011promoting}.
These implementations serve the role of the \emph{Monitor} in our framework. Depending on the specific context and use case, different ways to integrate usage tracking are required.

We do not try to solve the issue of tracking itself, as there are various proposals and concepts in related work.
Our transparency framework instead allows integrating \emph{Transparency by Design} into software systems.
If software is designed based on our framework, these technical ways to retroactively provide transparency over data usages are not required anymore.
Until then, they serve as important steps towards our goal.

\subsection{Instantiating the Framework}
\label{sec:transparency-instantiation}

Our transparency framework concept represents a software design pattern to enable \emph{Transparency by Design}.
It can be instantiated in various ways, depending on the specific context.
To verify the applicability of our concept, we instantiate it for a real-world use case provided by our industry partner.
Based on two application examples, we describe the steps and estimate the potential effort for integrating the framework into software systems.

As deliberated above, certain trust antecedents depend on the specific instantiation, specifically those related to the presentation (see Section~\ref{sec:concept-trustworthiness}). Accordingly, we describe how we systematically target those with our instantiated \emph{Display} (see Section~\ref{sec:clotilde}).

\subsubsection{Use Case}

We model the following use case after the real-world scenario provided by our company partner (a software development company with about $5000$ employees and US\$ $1$ bn. yearly revenue), aiming to make it as realistic as possible.

Consider data usage processes in internal company infrastructures. Data about employees are stored in various systems and accessed through a multitude of tools. Processes depend on the availability of data, but some usages of data are prohibited by a company agreement to protect employees.
Various data sources can be imagined.
Mirroring the real-world scenario provided by our industry partner, we consider as data sources an issue tracking software, a team messenger, and a version control system.

\subsubsection{Technical Overview}
\label{sec:instantiation-technical-overview}

In order to allow for an easy and step-by-step migration to the transparency framework for existing data usage processes, we instantiate the components of the framework as web services. These services can be deployed on internal company servers. Based on our concept (see Figure~\ref{fig:framework-concept}), we develop two services: The safekeeping service \emph{Overseer}, tasked with verifying and storing usage information, as well as the web front-end \emph{Clotilde UI}, which makes the usage information transparent to data owners. The tracking of data usages---represented by the \emph{Monitor} in our concept---is integrated into the data sources or connected analysis tools by way of a lightweight library.
Access to these tools is secured with logins that are stored in a state-of-the-art sign-on server, representing a company single sign-on system commonly employed in practice.
Combined, we refer to the instantiated components of the framework as the \emph{transparency toolchain}.

For each data source, a set of exemplary analyses is implemented and integrated into the toolchain, serving as potential usage paths for the data. This allows us to evaluate both the efficacy of the monitoring as well as the effect of the added transparency for different classes of software tools.

In our implementation, the company has control over the servers on which the toolchain is deployed, meaning it needs to be trusted to a certain degree.
One way to alleviate this issue is to replace the \emph{Overseer} storage with one that provides strong security guarantees (see, e.g.~\cite{accorsi2010bbox, ge2019permission}).
Our implementation therefore supports switching out the storage provider flexibly (see Section~\ref{sec:safekeeper-overseer}).

The code for all components of the transparency toolchain and exemplary implementations of the \emph{Monitor} is available on GitHub\footnote{\url{https://github.com/tum-i4/inverse-transparency}} and licensed with the permissive MIT open source license.\footnote{\url{https://opensource.org/licenses/MIT}}
We want to allow anyone to scrutinize it and, of course, utilize it for their own instantiation of the transparency framework.

\subsubsection{Trustworthiness: Presentation}
\label{sec:instantiation-trustworthiness}

As we have hinted at in Section~\ref{sec:concept}, many trust antecedents can be influenced directly by the presentation, meaning they depend largely on the instantiation chosen.
Accordingly, we discuss in more detail how we systematically foster trust by targeting those trust antecedents with our presentation.
We target \emph{subjective beauty}~\cite{hassenzahl2008aesthetics, yuksel2017brains}, \emph{feedback}~\cite{tognazzini2014first, hoff2015trust, bigras2018ai}, as well as \emph{usability}~\cite{fisher2008usability, faisal2016web, lee2015antecedents}.
To illustrate our following elaborations, find screenshots of the interface attached in Appendix~\ref{sec:appendix-user-interface}.

As the name suggests, \emph{subjective beauty} is highly subjective.
Inspired by previous work (see, e.g., \cite{hassenzahl2008aesthetics, yuksel2017brains}), we take the following steps:
We employ a symmetric layout with reduced visual complexity.
In addition, a graph displayed centrally and large numbers in the summary add visual variety.

To improve \emph{feedback}, we follow \cite{tognazzini2014first, hoff2015trust, bigras2018ai} and try to make feedback more accurate and explicit.
Clear and expressive wording is therefore employed in pop up messages and help texts.

Finally, to improve \emph{usability}, we follow \cite{fisher2008usability, faisal2016web, lee2015antecedents}.
We target seamless interactivity with no reloads.
The navigation is reduced to scrolling on a single page, reducing friction.
Finally, a consistent color layout and modern typography target user satisfaction.

Combined, these facets foster trust by improving central trust antecedents related to presentation~\cite{lee2004trust, hoff2015trust}.

\subsubsection{Display: Web App Clotilde UI}
\label{sec:clotilde}

Arguably the most important component of the framework is the display app \emph{Clotilde UI}, a single-page web application and the front-end of the toolchain. Even though most of the technical processes happen in the back-end, the users interact solely with the front-end, making it their de facto interface to the toolchain.
As we have discussed above, we target important trust antecedents related to presentation with this app.
It is implemented in JavaScript utilizing Svelte.\footnote{\url{https://svelte.dev}}
Supporting interactivity, ease of use and navigation, it is a single-page web application with no intermediate reloads.
Apart from this, we followed the steps outlined in Section~\ref{sec:instantiation-trustworthiness} to improve user experience and foster trustworthiness.

Conceptualization and implementation work spanned over seven months, with an approximate effort of 45 working days shared by two junior developers.
In total, it consists of 807 source lines of code (SLOC).

\subsubsection{Safekeeper: Web Service Overseer}
\label{sec:safekeeper-overseer}

The \emph{Overseer} service is tasked with verifying and storing usage information, and making those data available to \emph{Clotilde UI} to display. In that sense, it can be thought of as a back-end of the toolchain. We implemented it in Python. 
The storage is attached based on the \emph{data access object} pattern, which means it can be flexibly switched out according to the requirements. To ease set-up, we utilize an SQLite\footnote{\url{https://sqlite.org/}} database.

The modeling and development work spanned over seven months, with an approximate effort of 32 working days shared by two junior developers.
In total, it is composed of 565 SLOC.

\subsubsection{Monitor Instantiation: Jira Software Example}
\label{sec:instantiation-jira}

The \emph{Monitor} component of our transparency framework stands in for any data usage tracking.
Depending on the context, various options exist to enable such usage tracking (see Section~\ref{sec:concept-related-work}).
For existing software, one possibility is to add tracking functionality via a plugin.
In the following, we describe a possible implementation as plugins for a software tool. As an example, we look at Jira Software.

Jira Software is an issue tracking system. Issues represent work items and can be created by and assigned to anyone. Various data can be attached, such as a \emph{description}, a fixed-choice \emph{priority}, and \emph{links} to other issues. In addition, collaborators can also comment on issues, independently of who they are assigned to.

To access these data, multiple possibilities exist. Commonly, so-called \emph{dashboard plugins} are utilized. These allow analyses over, e.g., issues, people, or time frames. For our scenario, we developed four dashboard plugins covering common analyses. A representative example is the \emph{expert search}. Based on a use case from our industry partner, we implemented a search feature for colleagues by skill set. Without the added protection of the transparency framework, such a feature would not be allowed at the industry partner, as it might enable profiling if unchecked.
Similarly, the three other exemplary plugins also offer functionality that may be helpful but potentially enables misusage or discrimination. The plugins are detailed in Appendix~\ref{sec:appendix-plugins}.
All four plugins were developed following the principle of \emph{Transparency by Design}. That means that we already integrated a \emph{Monitor} into each plugin during development.

\begin{figure}[htbp]
	\centering
	\includegraphics[width=0.9\linewidth]{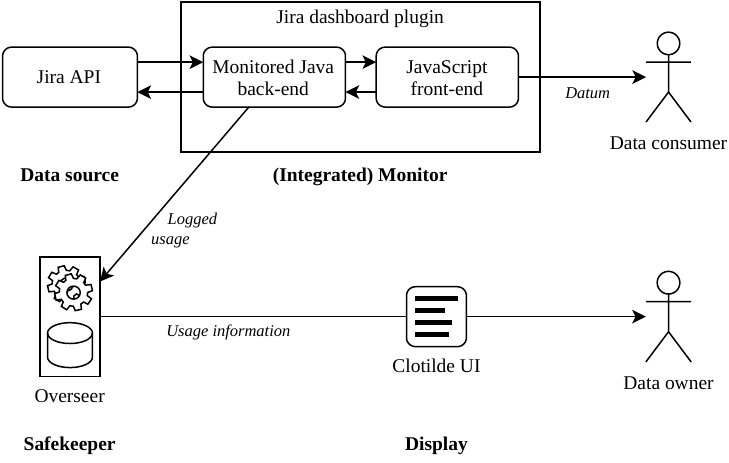}
	\caption{Exemplary instantiation of a \emph{Monitor} as part of a Jira dashboard plugin. Similar architectures can be employed for other tools with plugin capabilities. The toolchain components \emph{Overseer} and \emph{Clotilde UI} fulfill the \emph{Safekeeper} and \emph{Display} roles of the transparency framework concept.}
	\label{fig:framework-instantiation-jira}
\end{figure}

In practice, the dashboard plugins all follow the same basic architecture (see Figure~\ref{fig:framework-instantiation-jira}). This architecture is based on our transparency framework concept. Whenever data are requested from the Jira API, the Java back-end reports the usage to the \emph{Overseer} server, thereby enabling transparency into the data usage process.

The monitoring component is implemented in one self-contained Java class that is shared between all dashboard plugins, consisting of just 154 SLOC. The estimated development effort was just one working day for a junior developer.

\subsubsection{Monitor Instantiation: Standalone Analysis Tools}
\label{sec:instantiation-standalone}

Instead of integrating the monitors in plugins for an existing software tool, new tools can be developed with a monitor component already built in. This would most closely represent our vision of \emph{Transparency by Design}. For the standalone analysis tools, we followed this idea throughout development.
We implemented external analysis tools for the team messenger and the version control system from our use case.
Both tools were developed using Python, with the \emph{Monitor} integrated as a Python module. This code can be shared between both analysis tools, so it only needed to be developed once.

The monitoring module consists of just 52 SLOC, with an estimated development effort of less than half a working day for a junior developer. In addition to it being published to GitHub, we also distribute the module via the Python package index. 
With this, we aim to enable anyone to employ \emph{Transparency by Design} during development of Python-based analytics software.
Furthermore, the code can serve as an example for other \emph{Monitor} instantiations.

\subsection{Evaluation}
\label{sec:evaluation}

Our goal is to enable trustworthy transparency into data usage processes.
In order for our transparency framework to enable \emph{data sovereignty}, it depends on the acceptance and trust of individuals.
Therefore, we systematically developed the toolchain---an instantiation of our transparency framework---to foster trust and improve user experience.
In the following, we evaluate to what extent we managed to achieve this goal with the transparency toolchain.

We evaluate four central research questions:
First, was the toolchain trusted by participants of our study?
Second, did they develop personal attachment or affinity to the toolchain?
Third, did participants consider the toolchain to be usable and understandable?
And finally, was the transparency over data usages considered helpful?

An important factor of \emph{data sovereignty} is for individuals to be able to benefit from the usage of their data, as we have deliberated above. Without sensible use cases for data (see, e.g.,~\cite{waber2013people, gal2020breaking}), giving access to it may not make sense for them.
In our evaluation, we focus on the transparency framework as an enabler of trust.
In a follow-up case study at our industry partner, we want to incorporate the incentives for individuals to give access to their data.

As we take the first step towards implementing a completely new approach to data privacy, our evaluation is focused on assessing in-depth the effect this transparency has on individuals and how they experience it.
This helps to gather experiences and prepare case studies at our industry partner.
Accordingly, it is designed as a qualitative focus group study.

\subsubsection{Study Design}

To ground our results in reality, we emulated the real-world use case provided by our industry partner as closely as possible.
The tools we looked at (see Section~\ref{sec:transparency-instantiation}) and the process we followed therefore match this use case.

Our evaluation is a qualitative focus group study.
To generate realistic data and experiences, we created a university practicum running over three months.
Twelve master's students in computer science were tasked to conduct work in a setting closely mirroring the real-world scenario over two months' time. In the third month, they used analytics tools to analyze the personal data collected in the months prior. This allowed them to experience data usage both from the perspective of a data consumer as well as a data owner.
All analytics tools were integrated into our instantiated transparency framework, which meant that all usages of data were tracked and made available to data owners.
Through the web interface \emph{Clotilde UI}, the data owners had direct transparency into all data usages concerning their own data.

We employ a mixed-method evaluation design~\cite{wakeling2015integrating, garciagathright2018mixed}.
During the process, regular qualitative evaluations in the form of interviews and written self-reflections were conducted. In discussion sessions, the participants were asked to actively weigh their experience with the toolchain. In addition, the user experience and trustworthiness of the toolchain were evaluated with a questionnaire. The questions are derived from the state of the art in empirical evaluation of user trust and experience (see Section~\ref{sec:questionnaire-construction}).
Underlying the questions is a reflective measurement model~\cite{mackenzie2011construct}.
Each latent variable is targeted twice with a slightly reworded question to verify internal consistency.
Responses to the questionnaire were recorded on five-point Likert scales~\cite{likert1932technique, allen2007likert}.
In related work, both five-point and seven-point Likert scales are common (see, e.g., \cite{faisal2016web, hammer2015trust, merritt2011affective}).
The five-point scale was chosen over a seven-point scale to reduce complexity in the interpretation of the answers.
The uneven-numbered scale leads to the availability of a neutral option, which gives additional meaning to answers leaning to either side. If someone chooses a non-neutral option, their choice is explicitly in contrast to a neutral answer and not enforced by an even-numbered scale.

\subsubsection{Questionnaire Construction}
\label{sec:questionnaire-construction}

We derive the central constructs we target from the state of the art in empirical evaluation of user trust and experience.
For each latent variable, we surveyed existing work on evaluation instruments targeting the same constructs.

The latent variables we consider are \emph{trust}~\cite{corritore2005measuring, merritt2011affective, hammer2015trust}, \emph{perceived reliability}~\cite{madsen2000measuring, merritt2011affective}, \emph{personal attachment}, \emph{affinity}~\cite{madsen2000measuring, chien2014towards}, \emph{usability}~\cite{corritore2005measuring, roy2001impact}, and \emph{understandability}~\cite{komiak2006effects, faisal2016web}.

\subsubsection{Threats to Validity}

It is important to note the threats to the validity of this evaluation.
Firstly, the scenario was artificial and occurred within the university context, as it was not possible to conduct this in-depth evaluation at our industry partner.
This also means that participants had different stakes compared to real-world employees and were closer to the research and each other.
To remedy those issues, the scenario is modeled closely after the real-world use case.
In addition, we utilize a mixed-method design to enhance the informative value of the results. Besides the questionnaire results, we discuss extracts from the interviews and written self-reflections to help explain the results.

The utilized analytics tools were developed by participants themselves.
This reduced the potential effect of interpersonal trust (see Section~\ref{sec:concept-trustworthiness}) on their trust in the toolchain.

Finally, the number of participants ($n=12$) is comparatively low.
Our evaluation approach is a focus group study. In order to thoroughly weigh individual experiences and perspectives, a higher effort was required. This made it infeasible to include a larger number of participants.


\newcommand{\likertexplainer}{Participants ($n=12$) were asked to express their agreement on a five-point Likert scale (X-axis), from 1 for ``strongly disagree'' to 5 for ``strongly agree''.}

\subsubsection{RQ 1: Was the toolchain trusted?}

An important prerequisite for a discussion of further results is participants' underlying trust in the toolchain and their perception of its reliability as an important antecedent of trust~\cite{fogg1999elements, stanton2009effects, oleson2011antecedents}.
If this is not given, the validity of other results, but also the concept of providing trustworthy transparency on the whole, may be jeopardized.

Derived from the questionnaires in~\cite{madsen2000measuring, corritore2005measuring, merritt2011affective, everard2005presentation, hammer2015trust}, we defined the following statements:

\begin{enumerate}
	\setlength\itemsep{0em}
	\item The toolchain performs reliably.
	\item I feel that I could rely on the toolchain to function properly.
	\item I trusted the toolchain.
	\item I believe the toolchain to be trustworthy.
\end{enumerate}

\begin{figure}[htbp]
	\centering
	\includegraphics[width=0.9\linewidth]{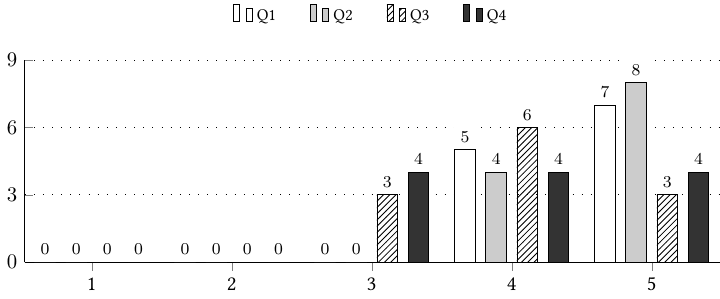}
	\caption{Results for the questions targeting trust and reliability (Q1--Q4). \likertexplainer}
	\label{fig:plot-trust-reliability}
\end{figure}

Following~\cite{gulati2019towards, chien2014towards, madsen2000measuring}, we asked participants for their level of agreement with each statement on a five-point Likert scale.
The results (see Figure~\ref{fig:plot-trust-reliability}) are very positive overall, with only three, respectively four, respondents choosing the neutral answer for the questions of trust (Q3, Q4).

As mentioned above, these results are by no means complete, nor do they offer a comprehensive picture. What they can confirm in this context is that this specific group of participants believed the toolchain to be trustworthy and reliable, in effect trusting its efficacy. That increases our confidence in the validity of the results of our other research questions.

\subsubsection{RQ 2: Did participants develop personal attachment or affinity to the toolchain?}

Derived from the questionnaires in~\cite{madsen2000measuring, chien2014towards}, we defined the following statements:

\begin{enumerate}
	\setlength\itemsep{0em}
	\setcounter{enumi}{4}
	\item I would have felt a sense of loss if the toolchain was unavailable and I could no longer use it.
	\item I felt a sense of attachment to using the toolchain.
	\item I found the toolchain suitable for me.
	\item I liked using the toolchain.
\end{enumerate}

\begin{figure}[htbp]
	\centering
	\includegraphics[width=0.9\linewidth]{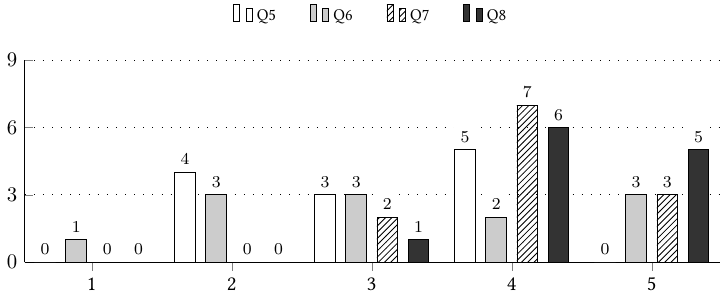}
	\caption{Results for the questions targeting personal attachment (Q5--Q8). \likertexplainer}
	\label{fig:plot-personal-attachment}
\end{figure}

Participants were asked to declare how much they agree with each statement.
The results (see Figure~\ref{fig:plot-personal-attachment}) show a trend towards the values 4 and 5, representing agreement with the posed statements. The questions of personal attachment (Q5, Q6) show the most variety in the answers, with participants almost evenly divided for both. On the other hand, the questions targeting personal affinity (Q7, Q8) are answered positively almost unanimously.

When looking into the results of our interviews and participants' reflections, we can find possible explanations for this result. One participant explained: ``I haven't used [the transparency toolchain] much. Therefore I wouldn't really miss it. I don't think a piece of software will ever give a sense of attachment.'' Another argued: ``I really liked the experience and thing [sic] this would be quite useful for a couple of use cases, however, I do not think I that I would suddenly miss the toolchain if it were gone.'' Finally, one reasoned: ``The circumstances were not critical enough to feel a loss, might be different in a corporate environment. Still I think it is suitable and I liked using it a lot.''

These examples may help explain some of the questionnaire responses. It seems that due to the semi-artificial nature of the scenario, the transparency provided by the toolchain was not judged to be as crucial by some participants.
Still, the positive outcome regarding personal affinity is promising.

\subsubsection{RQ 3: Did participants consider the toolchain to be usable and understandable?}

Importantly, we want to assess the perceived usability of the transparency provided by the toolchain.
Derived from the questionnaires in~\cite{corritore2005measuring, roy2001impact, komiak2006effects, faisal2016web}, we defined the following statements and asked participants for their level of agreement:

\begin{enumerate}
	\setlength\itemsep{0em}
	\setcounter{enumi}{8}
	\item I found the system easy to use.
	\item I could find easily what I was looking for on the interface.
	\item I am familiar with how the toolchain works.
	\item I found the interface easy to learn to use.
\end{enumerate}

As mentioned above, underlying all of our questions is a reflective measurement model. With these questions, we can see an increased fuzziness which latent variable is measured. The main aspects that overlap here are the interface itself and the information represented \emph{through} the interface. Accordingly, we again follow up the questionnaire responses with a discussion of the results from our interviews and participants' self-reflections below.

\begin{figure}[htbp]
	\centering
	\includegraphics[width=0.9\linewidth]{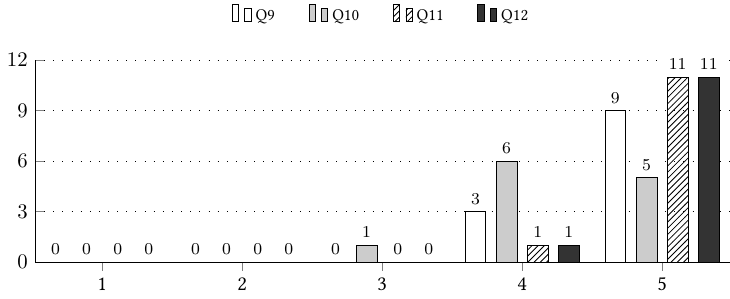}
	\caption{Results for the questions targeting usability and understandability (Q9--Q12). \likertexplainer}
	\label{fig:plot-usability-understandability}
\end{figure}

The results (see Figure~\ref{fig:plot-usability-understandability}) are very positive, with respondents almost unanimously agreeing with the statements targeting understandability, with just somewhat reduced levels of agreement for statements targeting usability. To address the fuzziness as to which latent variable these answers target, we discuss some related quotes from the interviews and reflections by participants in detail below.

As we assumed, many mentioned the user interface and not the represented information.
This validates our efforts to improve user experience and usability to foster trust.
Responses include: ``I liked the very clean and intuitive structuring of \emph{Clotilde},'' ``The \emph{Clotilde UI} feels very user-friendly and easy to use,'' ``\emph{Clotilde} has a nice user interface,'' among others. \emph{Clotilde UI} is the user interface of the toolchain (see Section~\ref{sec:clotilde}).

Others discuss the information that they are given access to (the transparency provided). All of the responses directly targeting this aspect are positive. They include: ``I enjoyed investigating who analysed (me) the most. It seems a good approach [...]'', ``It was interesting to see that most of the requests were accessing Jira data'', and ``We can get a lot of information at once and [...] the information is still clear to the user.''
These statements hint at the usefulness of the provided transparency.

Other statements indicated the limitations of our approach, which we will deliberate in the following research question.

\subsubsection{RQ 4: Was the transparency over data usages helpful?}

To highlight and discuss the overall helpfulness of our approach of providing transparency into data usage processes, we focus on individual results of our qualitative evaluation. In the following, we highlight statements and reflections that offer an understanding of how helpful the approach was seen in general, and where specific limitations of our instantiation lie.
The quotes are parts of interviews and long-form written self-reflections that were regularly submitted by participants.

We received various thoughtful comments regarding this question. We do not try to summarize, but instead highlight some exemplary quotes. This is not by any means complete, but may help understand the individual perspectives.

Some responses hint at the limitations of our evaluation setting. Due to the semi-artificial nature of the setting, some considered the value compared to a real-world scenario to be reduced, saying that ``it will be more valuable for me in a company context'' or ``The circumstances were not critical enough [...], might be different in a corporate environment.''

With those limitations in mind, the reflections and discussions were very fruitful. One argued that \emph{Transparency by Design} can be considered a ``bureaucratic approach'', which could be both an advantage and a disadvantage. They argued that, as the concept is to integrate into the infrastructure, circumventing the transparency framework for malicious purposes becomes a challenge. On the other hand, they consider the problem of this ``bureaucratic'' nature making the underlying issues seem almost ``boring'' or ``tedious'' if the volume were to rise too much, comparing them to cookie notices found on websites.
Some of the other participants' accounts confirm this to be a problem already in our relatively short evaluation period, which covered just three months. The most concise summary of this issue is: ``There were constantly more than 100 weekly accesses to my data. Therefore, I lost the overview of the requests quite early and did not investigate them in detail.''
This is a valuable point that we have pondered as well. We think that this problem can be circumvented in various ways. One of them could be the automatic analysis of logged usage information to try to find anomalies or potentially malicious usages, for example with machine learning approaches. This has the potential to reduce the mental load for data owners by reducing the amount of information they are presented with.
We discuss this and other potential ways to address limitations of our approach in the discussion (see Section~\ref{sec:discussion}).

Ultimately, multiple participants mentioned experiencing a feeling of increased control and safety, a central aspect to achieve \emph{data sovereignty}. Examples include: ``I [...] felt more secure about the use of my data'' or ``I feel safer as no one can play with [my data] without my awareness''.
This gives us confidence that our approach aims in the right direction and has the potential to enable \emph{data sovereignty} for individuals.

\subsection{Results}

The goal of our transparency framework is to enable \emph{Transparency by Design}, in order to facilitate the \emph{data sovereignty} of individuals.
To achieve this goal, the framework needs to fulfill two main requirements:
Instantiating it needs to be intuitive and easy (1) and users need to consider the provided transparency to be trustworthy and empowering (2).

In our instantiation (see Section~\ref{sec:transparency-instantiation}), we describe the changes to the software design that enable \emph{Transparency by Design} based on two exemplary real-world scenarios.
Even for junior developers, the effort required is minor, with \emph{Monitor} implementations completed in less than one working day.
Furthermore, all of our code is released as open source and developers can base their work off of it. This further reduces the steps required to enable transparency into data usage processes.
Therefore, requirement (1) is met.

The evaluation (see Section~\ref{sec:evaluation}) follows a focus group over three months to analyze in-depth the effect this added transparency can have.
Prerequisites for the validity of our results are shown to be given, in that participants trust the toolchain to reliably provide the promised transparency.
In further research questions, we find that participants do not report increased personal attachment to the toolchain.
This could be explained by the semi-artificial nature of the setting.
Usability and understandability on the other hand are ranked very high. Quotes from participants' reflections show that this is mainly due to the intuitive user interface.
Finally, there is evidence in the evaluation results that the provided transparency is considered helpful and empowering.
Multiple participants specifically highlighted that they experienced a feeling of increased control and safety.
Accordingly, requirement (2) is met as well.

\section{Discussion}
\label{sec:discussion}


There are some limitations to our approach of enabling full transparency into data usage processes.
First, the potential of our transparency framework depends in large parts on the security guarantees that can be given.
If the framework logging can be easily circumvented, the tracking of usages becomes a feeble effort.
As soon as data are given away, our control over it ends (see, e.g., \cite{pretschner2008usage}), requiring advanced technical protections (e.g.,~\cite{ahmadvand2019taxonomy, costan2016intel}) that might be infeasible.
This is why we envision every software tool to integrate \emph{Transparency by Design}.
Yet, realistically, such comprehensive integration of transparency may only be implementable in semi-closed environments such as internal company infrastructures or within governmental institutions.

Furthermore, as we found in our evaluation, the volume of data usages will most probably be significantly too large for individuals to sensibly oversee every individual usage.
There is a risk of users overlooking relevant notifications if they get too many benign ones, due to generalization or habituation~\cite{vance2019fog}.
Potentially, automatically analyzing tracked data usages could remedy this.
Techniques such as those utilized in anomaly and intrusion detection (see, e.g.,~\cite[p.~5]{chandola2009anomaly} or \cite[p.~9]{zieglmeier2018real}) could be applied to detect deviations from expected usage patterns.
Data owners would then be provided with a summary of usages that are deemed to be noteworthy or problematic.
Such periodic summaries or more irregular notifications may be preferable to users, depending on the type of data~\cite{naeini2017privacy}.
This might also have the potential to reduce their mental load when examining the recorded usages.

We also see the issue of missing interpretation and contextualization of the information that users get access to through our transparency framework.
Multiple participants of our focus group expressed such concerns when accessing data.
They worried that their recorded data usages could be misinterpreted by data owners.
Some asked for an opportunity to explain their usages, which can be a remedy in cases of, e.g., sensitive data or unusual usage patterns.
Potentially, this could also happen retroactively, with data owners being able to request justification for individual usages through the transparency framework.

This highlights a second-order effect of the provided transparency: data consumers' tracked usage information represents personal data about them.
As previous works show (e.g.,~\cite{demontjoye2014openpds}), metadata themselves need to be protected, as they contain sensitive information.
Therefore, we consider it important to incorporate technical protections (see, e.g.,~\cite{deng2011privacy}) for tracked usage data.

In cases where data are used by algorithmic systems, making individual data usages transparent is not enough.
These systems can be black boxes even to their developers, meaning that individuals with no technical background would lack understanding of the meaning of individual data usages.
Existing work on algorithmic accountability and explainable AI (see~\cite{wieringa2020account, liao2020questioning}) could be integrated to give justifications for these data usages.
In our view, this would be a comprehensive solution for automatic decision making.
Through the transparency framework, an understandable explanation for algorithmic decisions (see, e.g.,~\cite{mittelstadt2019explaining, miller2019explanation}) could be provided to individuals.


Complete understanding of all data usages or perfect oversight by data owners may not be necessary, though.
Potentially, the most powerful effect of providing transparency into data usage processes might be to \emph{deter} misusage.
If just a slight chance of detection and appropriate consequences exists (for inspiration, see \cite[p.~36]{mundie2014privacy}), data consumers might be discouraged to misuse data they have access to, giving back power and control to data owners.
\citeauthor{hall2017accountability} describe this as \emph{felt accountability}~\cite{hall2017accountability}.
We plan to investigate this potential effect in a case study. In addition, we want to take the next step towards \emph{data sovereignty}: Allowing data owners to express in an intuitive and usable way which usages of data they want to allow, and which not.

\section{Conclusion}

In this paper, we describe how to address individuals' lack of oversight over technical systems processing their data, which can lead to discrimination and hidden biases that are hard to uncover.
Based on previous conceptual work, we develop a transparency framework that allows integrating transparency into software systems by design.
In our view, this oversight can enable individuals' \emph{data sovereignty}, allowing them to benefit from sensible usage of their data while addressing the potentially harmful consequences of data misusage.
To foster user trust and acceptance, we take systematic steps to improve the trustworthiness of the framework, following insights from the state of the art in user trust and experience research.
Our deliberations allow system designers to incorporate these results in the designs of their systems.
Finally, we instantiate and evaluate the transparency framework in a qualitative focus group study to assess its impact on individuals.
We find that it satisfies usability and trustworthiness requirements.
The provided transparency was experienced as beneficial and participants felt empowered by it.
Therefore, we conclude that our transparency framework enables designing software with \emph{Trustworthy Transparency by Design}.

\section*{Acknowledgments}

We thank Antonia Maria Lehene for her help in creating and validating our user trust questionnaire.

This work was supported by the Federal Ministry of Education and Research (BMBF) under grant no. 5091121.

\bibliographystyle{plainnat}
\bibliography{references}

\clearpage
\appendix

\section{User Interface Screenshot}
\label{sec:appendix-user-interface}

\begin{figure}[h!]
	\centering
	\includegraphics[width=\linewidth]{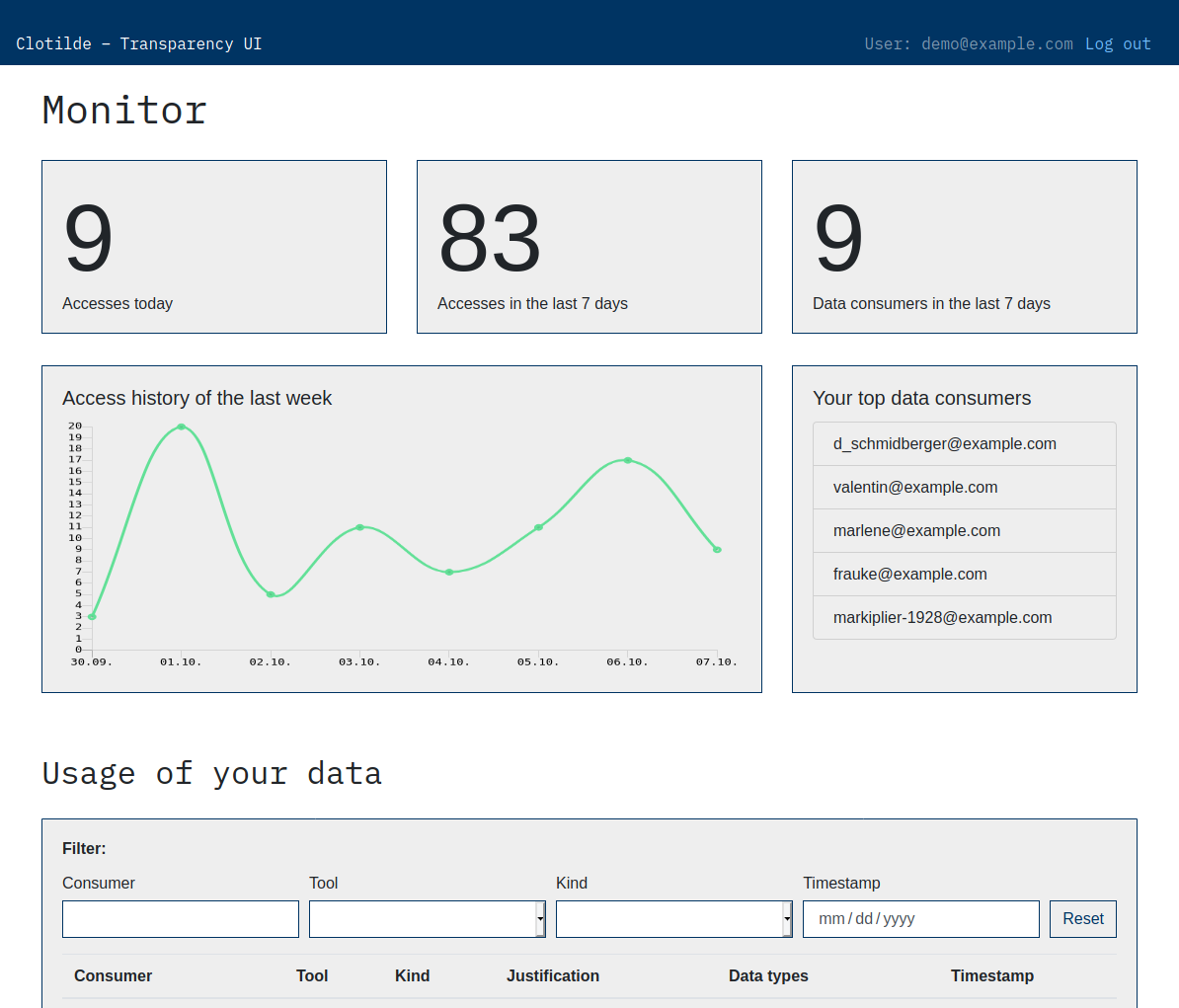}
	\caption{The user interface of \emph{Clotilde}. Large numbers and a graph add visual variety. The symmetric layout reduces visual complexity and can improve \emph{subjective beauty}. It is a single-page app, meaning navigation is simple.}
	\label{fig:clotilde-screenshot}
\end{figure}

\section{Analysis Plugins}
\label{sec:appendix-plugins}

\begin{table}[h!]
	\begin{tabular}{ l p{5.14cm} }
		\toprule
		\textbf{Plugin} & \textbf{Description} \\
		\midrule
		Expert search   & Find colleagues based on the tools and technologies they work on. \\
		Supporter list  & List those developers that review most issues. \\
		Leaderboard     & Rank all developers based on their approximated productivity, calculated from completed issue count, time estimate, and priority. \\
		Who needs help? & Show the users with most uncompleted issues.\\
		\bottomrule
	\end{tabular}
	\caption{The exemplary dashboard plugins developed for Jira. Each plugin represents a useful scenario, but potential for misusage.}
	\label{tab:dashboard-plugins}
\end{table}

\end{document}